\documentclass[modern]{aastex63}

\newcommand\gaia{\emph{Gaia}}
\begin{document}

\title{\gaia\ pulsars and where to find them in EDR3}

\correspondingauthor{John Antoniadis}
\email{jantoniadis@ia.forth.gr}

\author[0000-0003-4453-3776]{John Antoniadis}
\affiliation{Max-Planck-Institut f\"ur Radioastronomie, Auf dem H\"ugel 69, 53121 Bonn, Germany}
\affiliation{Argelander-Institut f\"ur Astronomie, Universit\"at Bonn, Auf dem H\"ugel 71, 53121 Bonn, Germany}
\affiliation{Institute of Astrophysics, Foundation for Research and Technology-Hellas, Voutes, 71110 Heraklion, Greece}

\begin{abstract}
The Early \gaia\  Data Release 3 (EDR3) provides precise astrometry for nearly 1.5 billion sources across the entire sky. A few tens of these are associated with neutron stars in the Milky Way and Magellanic Clouds.   
Here, we report on a search for EDR3 counterparts to known rotation-powered pulsars using the  method outlined in \cite{antoniadis2020c}. 
A cross-correlation between  EDR3 and the ATNF pulsar catalogue identifies 41 close astrometric pairs ($\lesssim 0\farcs5$ at the  reference epoch of the pulsar position). Twenty six of these are related to previously-known optical counterparts, while the rest are candidate pairs that require further follow-up. 
Highlights include the Crab Pulsar (PSR\,B0531+21), for which EDR3 yields a distance of $2.08^{+0.78}_{-0.45}$\,kpc (or $2.00_{-0.38}^{+0.56}$\,kpc taking into account the dispersion-measure prior;  errors indicate 95\% confidence limits)  and PSR\,1638$-$4608, a pulsar thus-far considered to be isolated that lies within 0\farcs056 of a \gaia\ source.    
\end{abstract}

\section{Introduction}
Most of the $\sim$2900 known pulsars are too faint to be detected at optical 
wavelengths \citep{mignani2014}. However, $\sim 100$ have {(sub-)}stellar 
binary companions bright enough to be studied with  space-borne and 
ground-based optical telescopes. These systems provide insights into  important physical processes, from stellar evolution and accretion, to the dynamics of 
curved spacetime \citep{Wex:2020ald}. In a recent work \citep{antoniadis2020c} 
We used the second \gaia\ data release \citep[DR2][]{Lindegren:2018cgr} to 
perform a systematic search for optical counterparts to 1534 rotation-powered 
pulsars. This search identified 22 pulsars with 
previously known counterparts and 8 additional candidate companions to young 
pulsars. 
Here we update the results in \cite{antoniadis2020c} using the  recent \gaia\ 
EDR3  \citep{brown2020,lindegren2020} and Version 
1.64\footnote{\url{https://www.atnf.csiro.au/research/pulsar/psrcat/}; Accessed on December 6, 2020} of the ATNF Pulsar catalogue \citep[PSRCat;][]{Manchester:2004bp}.

\section{Results}
PSRCat v1.64 contains 1670 pulsars with positions known to better than $0\farcs5$. As in \cite{antoniadis2020c}, potential EDR3 counterparts to these sources were identified by propagating \gaia\ astrometric solutions back to the reference epoch of each pulsar's position (\texttt{PosEpoch} in PSRCat).  As a rule,  sources with angular separations larger than the $2\sigma$ positional error were discarded. The latter was 
 calculated by taking into account the positional uncertainties of the 
 pulsar and the corresponding EDR3 source (at \texttt{PosEpoch}). 
 An error of $0\farcs25$ was added in quadrature to 
 account for possible systematic uncertainties, e.g. possible rotations between
 the slightly different reference frames and errors due to strong optical variability  \citep[see][for a more detailed justification]{antoniadis2020c}.
However, for most positive matches the systematic uncertainties appear to be smaller. 

 Table\,\ref{tab:1} lists all close astrometric pairs with a probability 
 of chance coincidence smaller than 10\%. Given this threshold, the 
 number of false positives should be $\mathcal{O}(4)$. 
Most MSPs companions are located between the main sequence and the WD cooling branch.  There appears to be no obvious correlation between HR position and orbital period  \citep[for such a correlation in cataclysmic variables that occupy the same region see ][]{Abril:2019yfu}. 



\subsection{Notable sources}

\begin{itemize}
    \item {\bf PSR\,B0531+21} The distance to the Crab Pulsar and its nebula is a long-standing issue in astronomy \citep{Kaplan:2008qm}. \gaia\ EDR3 gives a trigonometric parallax of $\hat{\pi}_{\rm G}=0.51\pm0.08$\,mas for the source. This translates to a distance of  $2.08^{+0.78}_{-0.45}$\,kpc \citep[see][for details on distance priors]{antoniadis2020c}. Combining the latter with the distance estimate derived using the pulsar dispersion measure and the NE2001 model for the free-electron density in the Galaxy \citep{cordes2002}, one finds  $2.00_{-0.38}^{+0.56}$\,kpc (95\% C.L.). This is somewhat smaller than the DR2 distance estimate of $2.55_{-0.62}^{+0.79}$\,kpc. 
    
    \item{\bf PSR\,J1638$-$4608}  \gaia\,\texttt{5992089022760118400} was identified as a potential counterpart to this young pulsar in \cite{antoniadis2020c}. However, a full five-parameter astrometric solution for the source was reported for the first time  in EDR3. With the latter, the concurrent angular separation between the pulsar and the source becomes $0\farcs0562\pm 0\farcs004_{\rm stat.} \pm 0\farcs2500_{\rm sys}$. This makes a physical association between the two objects even more likely. The inferred luminosity and color of the source are consistent with a massive main sequence star. Assuming both sources are at the same distance ($3.95^{2.55}_{-2.05}$\,kpc, taking into account the DM distance prior), the projected angular separation would be $222^{+143}_{-116}$\,AU. While the timing data for the pulsar do provide evidence for a periodic modulation \citep{Kerr:2015xth,Parthasarathy:2019txt}, this is unlikely to be caused by a massive binary companion. 
\end{itemize}

\section{Conclusions}
Overall, there are only minor differences between DR2 and EDR3.  With the exception of the sources listed in Table\,\ref{tab:1} most rotation-powered pulsars do not appear to have an optical counterpart brighter than the  \gaia\ sensitivity threshold.
This implies a binary fraction of young pulsars $f_{\rm young}^{\rm 
true}\leq 5.2(8.2)\%$ under the realistic(conservative) assumptions for the binary 
properties and current sensitivity thresholds outlined in \cite{antoniadis2020c}. 

We recommend further radio timing follow up of the 15 new candidates in Table~\ref{tab:1} to investigate their relation with the coincident \gaia\ sources.

\begin{longrotatetable}
\begin{deluxetable}{lcccccccccccccrr}
\tablecaption{Pulsars is \gaia\ eDR3}\label{tab:1}
\tabletypesize{\scriptsize}
\tablehead{\colhead{Name} & 
\colhead{$\theta$} & 
\colhead{ $\sigma_{\rm pos}$} & 
\colhead{$\hat{\pi}_{\rm G}$} & 
\colhead{ $\hat{\pi}_{\rm r}$} &
\colhead{$\mu_{\alpha}$} & 
\colhead{ $\mu_{\delta}$} & 
\colhead{ $\mu_{\alpha,r}$} & 
\colhead{ $\mu_{\delta,r}$} & 
\colhead{ $P_s$} &  
\colhead{ P$_{\rm b}$}  & 
\colhead{DM} & 
\colhead{$m_{\rm g}$}& 
\colhead{P$_{\rm assoc}$} & 
\colhead{d} & 
\colhead{ $M_{\rm g}$} \\ 
\colhead{} & 
\colhead{ $\farcs$} & 
\colhead{ $\farcs$} & 
\colhead{mas} & 
\colhead{ mas} &
\colhead{mas\,yr$^{-1}$} & 
\colhead{mas\,yr$^{-1}$} & 
\colhead{ mas\,yr$^{-1}$} & 
\colhead{mas\,yr$^{-1}$} & 
\colhead{ s} &  
\colhead{days} & 
\colhead{mag} & 
\colhead{pc\,cm$^{-3}$} & 
\colhead{$m_{\rm g}$} &  
\colhead{kpc} & 
\colhead{mag} 
} 
\tablenum{1}
\startdata
J0045$-$7319 & $0.5043$ & $0.2689$ & $-0.04(4)$ & $-$ & $0.58(5)$ &  $-1.49(5)$ &  $-$ &  $-$ &  $0.926$ &  $51.17$ &  $105.4$ &  $16.20$ &  $0.91$ & $49.97^{+1.13}_{-1.13}$ & $-4.85^{+0.05}_{-0.05}$ \\
J0337+1715 & $0.0081$ & $0.2500$ & $0.54(20)$ &$-$ & $5.48(19)$ &  $-4.43(16)$ &  $-$ &  $-$ &  $0.003$ &  $1.63$ &  $21.3$ &  $18.05$ &  $1.00$ & $1.69^{+0.62}_{-0.47}$ & $6.59^{+0.68}_{-0.70}$ \\
J0348+0432 & $0.0021$ & $0.2508$ & $-0.04(78)$ &$-$ & $3.44(134)$ &  $-0.23(89)$ &  $4.04(16)$ &  $3.50(60)$ &  $0.039$ &  $0.10$ &  $40.5$ &  $20.59$ &  $1.00$ & $2.36^{+1.20}_{-1.03}$ & $7.94^{+0.90}_{-1.25}$ \\
J0437$-$4715 & $0.3166$ & $0.2501$ & $7.10(52)$ & $6.37(90)$ & $121.65(57)$ &  $-70.70(65)$ &  $121.4385(20)$ &  $-71.4754(20)$ &  $0.006$ &  $5.74$ &  $2.6$ &  $20.35$ &  $1.00$ & $0.1570^{+0.0030}_{-0.0040}$ & $14.37^{+0.04}_{-0.06}$ \\
J1012+5307 & $0.0117$ & $0.2500$ & $1.74(29)$ & $0.71(17)$ & $2.74(29)$ &  $-25.92(27)$ &  $2.61(1)$ &  $-25.48(1)$ &  $0.005$ &  $0.60$ &  $9.0$ &  $19.59$ &  $0.99$ & $1.73^{+1.63}_{-0.58}$ & $8.36^{+1.44}_{-0.89}$ \\
J1023+0038 & $0.0031$ & $0.2500$ & $0.69(7)$ & $0.73(2)$ & $4.61(7)$ &  $-17.28(8)$ &  $4.76(3)$ &  $-17.34(4)$ &  $0.002$ &  $0.20$ &  $14.3$ &  $16.23$ &  $0.99$ & $1.37^{+0.07}_{-0.07}$ & $5.48^{+0.11}_{-0.11}$ \\
J1024$-$0719 & $0.0753$ & $0.2501$ & $0.86(28)$ & $0.80(30)$ & $-35.46(32)$ &  $-48.35(36)$ &  $-$ &  $-$ &  $0.005$ & $-$&  $6.5$ &  $19.15$ &  $1.00$ & $1.87^{+3.22}_{-0.90}$ & $7.67^{+2.17}_{-1.42}$ \\
J1048+2339 & $0.0087$ & $0.2501$ & $0.49(44)$ &$-$ & $-15.45(35)$ &  $-11.62(34)$ &  $-18.70$ &  $-9.40$ &  $0.005$ &  $0.25$ &  $16.7$ &  $19.59$ &  $1.00$ & $2.20^{+1.14}_{-0.92}$ & $7.83^{+0.91}_{-1.18}$ \\
J1227$-$4853 & $0.0622$ & $0.2500$ & $0.46(13)$ & $-$ & $-18.77(11)$ &  $7.30(90)$ & $ -$ &  $-$ &  $0.002$ &  $0.29$ &  $43.4$ &  $18.07$ &  $0.99$ & $2.60^{+2.33}_{-0.92}$ & $5.86^{+1.52}_{-0.99}$ \\
B1259$-$63 & $0.0597$ & $0.2500$ & $0.44(1)$ &$-$ & $-7.090(10)$ &  $-0.340(10)$ &  $-6.60(18)$ &  $-4.40(14)$ &  $0.048$ &  $1236.72$ &  $146.7$ &  $9.63$ &  $0.96$ & $2.26^{+0.12}_{-0.11}$ & $-6.58^{+0.34}_{-0.32}$ \\
J1311$-$3430 & $0.0116$ & $0.2506$ & $1.93(97)$ & $-$ & $-6.1(16)$ &  $-5.14(68)$ &  $-$ & $-$ &  $0.003$ &  $0.07$ &  $37.8$ &  $20.44$ &  $1.00$ & $2.32^{+1.49}_{-1.33}$ & $8.56^{+1.12}_{-1.88}$ \\
J1417$-$4402 & $0.3816$ & $0.2805$ & $0.20(5)$ & $-$ & $-4.76(4)$ &  $-5.10(5)$ &  $-$ &  $-$ &  $0.003$ &  $5.37$ &  $55.0$ &  $15.77$ &  $0.99$ & $5.06^{+2.50}_{-1.35}$ & $2.02^{+0.98}_{-0.74}$ \\
J1431$-$4715 & $0.0471$ & $0.2500$ & $0.53(13)$ & $-$ & $-11.82(14)$ &  $-14.52(15)$ &  $-7.0(30)$ &  $-8.0(40)$ &  $0.002$ &  $0.45$ &  $59.4$ &  $17.73$ &  $0.99$ & $2.24^{+1.78}_{-0.72}$ & $5.84^{+1.38}_{-0.89}$ \\
J1723$-$2837 & $0.3120$ & $0.2500$ & $1.07(4)$ & $-$ & $-11.73(4)$ &  $-24.05(3)$ &  $-$ &  $-$ &  $0.002$ &  $0.62$ &  $19.7$ &  $15.54$ &  $0.93$ & $0.94^{+0.07}_{-0.06}$ & $4.17^{+0.16}_{-0.15}$ \\
J1810+1744 & $0.0592$ & $0.2505$ & $0.65(54)$ &$-$ & $7.54(45)$ &  $-4.19(51)$ &  $-$ &  $-$ &  $0.002$ &  $0.15$ &  $39.7$ &  $20.00$ &  $0.99$ & $2.42^{+1.38}_{-1.14}$ & $7.61^{+0.98}_{-1.51}$ \\
J1816+4510 & $0.0019$ & $0.2500$ & $0.22(10)$ & $-$ & $-0.06(12)$ &  $-4.40(12)$ &  $5.30(80)$ &  $-3.0(10)$ &  $0.003$ &  $0.36$ &  $38.9$ &  $18.20$ &  $1.00$ & $4.31^{+2.07}_{-1.44}$ & $4.95^{+0.85}_{-0.88}$ \\
J1957+2516 & $0.0078$ & $0.2505$ & $2.15(85)$ & $-$ & $-4.49(52)$ &  $-12.29(101)$ &  $-$ &  $-$ &  $0.004$ &  $0.24$ &  $44.1$ &  $20.28$ &  $0.96$ & $2.34^{+1.68}_{-1.48}$ & $5.90^{+2.23}_{-4.14}$ \\
B1957+20 & $0.0889$ & $0.2505$ & $1.2(14)$ & $-$ & $-19.3(13)$ &  $-26.6(13)$ &  $-16.00(50)$ &  $-25.80(60)$ &  $0.002$ &  $0.38$ &  $29.1$ &  $20.17$ &  $0.96$ & $1.88^{+1.03}_{-0.92}$ & $7.90^{+1.15}_{-1.71}$ \\
J2032+4127 & $0.0476$ & $0.2500$ & $0.57(2)$ &$-$ & $-2.88(2)$ &  $-0.95(2)$ &  $-$ &  $-$ &  $0.143$ &  $16835.00$ &  $114.7$ &  $11.28$ &  $0.99$ & $1.76^{+0.08}_{-0.08}$ & $-5.11^{+0.11}_{-0.11}$ \\
J2129$-$0429 & $1.3038$ & $0.2500$ & $0.48(7)$ & $-$ & $12.10(7)$ &  $10.19(6)$ &  $-$ &  $-$ &  $0.008$ &  $0.64$ &  $16.9$ &  $16.82$ &  $1.00$ & $2.18^{+0.74}_{-0.44}$ & $5.05^{+0.63}_{-0.49}$ \\
J2215+5135 & $0.0869$ & $0.2500$ & $0.30(23)$ & $-$ & $0.01(24)$ &  $2.24(24)$ &  $170.0(220)$ &  $82(24)$ &  $0.003$ &  $0.17$ &  $69.2$ &  $19.20$ &  $0.98$ & $2.99^{+1.49}_{-1.15}$ & $6.37^{+1.12}_{-1.20}$ \\
J2339$-$0533 & $0.1251$ & $0.2504$ & $0.53(18)$ & $-$ & $3.92(20)$ &  $-10.28(19)$ &  $11.00(4.00)$ &  $-29(10)$ &  $0.003$ &  $0.19$ & $-$ &  $18.79$ &  $1.00$ & $1.57^{+0.50}_{-0.39}$ & $7.77^{+0.61}_{-0.61}$ \\
\hline
\multicolumn{15}{c}{Single Pulsars} \\
\hline
B0531+21 & $0.0460$ & $0.2501$ & $0.51(8)$ & $-$ & $-11.51(10)$ &  $2.30(6)$ &  $-14.70(80)$ &  $2.00(80)$ &  $0.033$ & $-$ &  $56.8$ &  $16.53$ &  $1.00$ & $2.08^{+0.78}_{-0.45}$ & $3.71^{+0.71}_{-0.54}$ \\
B0540$-$69 & $0.0279$ & $0.2503$ & $0.6(13)$ & $-$ & $3.6(21)$ &  $5.0(23)$ &  $-$ &  $-$ &  $0.051$ & $-$ &  $146.5$ &  $20.77$ &  $0.93$ & $62.10^{+1.90}_{-1.90}$ & $-5.84^{+0.07}_{-0.07}$ \\
\hline
\multicolumn{15}{c}{Small angular separation but most likely not associated} \\
\hline
B1953+29 & $0.1399$ & $0.2500$ & $0.47(17)$ &$-$& $-3.17(14)$ &  $-7.83(17)$ &  $-$ &  $-$ &  $0.006$ &  $117.35$ &  $104.5$ &  $18.70$ &  $0.97$ & $3.54^{+3.23}_{-1.70}$ & $3.19^{+1.97}_{-2.28}$ \\
J1435$-$6100 & $0.4139$ & $0.2500$ & $0.42(25)$ &$-$ & $-5.26(22)$ &  $-3.23(24)$ &  $-$ &  $-$ &  $0.009$ &  $1.35$ &  $113.7$ &  $18.92$ &  $0.95$ & $2.84^{+1.55}_{-1.17}$ & $-1.11^{+5.17}_{-4.33}$ \\
\hline
\multicolumn{15}{c}{Candidates} \\
\hline
J0534$-$6703 & $0.5187$ & $0.2872$ & $0.16(16)$ &$-$ & $1.27(26)$ &  $0.44(19)$ &  $-$ &  $-$ &  $1.818$ &$ -$ &  $94.7$ &  $18.86$ &  $0.94$ & $62.10^{+1.90}_{-1.90}$ & $-0.64^{+0.07}_{-0.07}$ \\
J1437$-$6146 & $0.4985$ & $0.2805$ & $0.31(0.03)$ & $-$ & $-5.00(0.03)$ &  $-2.95(0.03)$ &  $-$ &  $-$ &  $0.468$ & $-$ &  $200.5$ &  $15.68$ &  $0.90$ & $3.31^{+0.73}_{-0.51}$ & $-0.02^{+1.12}_{-0.84}$ \\
J1509$-$6015 & $0.1999$ & $0.2872$ & $0.19(11)$ & $-$ & $-5.25(11)$ &  $-2.39(11)$ &  $-$ &  $-$ &  $0.339$ & $-$ &  $423.6$ &  $17.76$ &  $0.91$ & $4.88^{+4.05}_{-1.96}$ & $0.42^{+4.55}_{-2.68}$ \\
J1542$-$5133 & $0.3258$ & $0.3103$ & $-0.21(27)$ & $-$ & $-2.25(34)$ &  $-4.86(28)$ &  $-$ &  $-$ &  $1.784$ & $-$ &  $186.0$ &  $19.03$ &  $0.91$ & $5.37^{+4.91}_{-2.64}$ & $2.81^{+3.76}_{-2.72}$ \\
J1546$-$5302 & $0.5576$ & $0.3094$ & $-$ & $-$ &$-$ &  $-$ &  $-$ & $-$ &  $0.581$ &  - &  $287.0$ &  $21.11$ &  $0.97$ & $5.33^{+1.07}_{-1.07}$ & $-5.14^{+2.92}_{-3.01}$ \\
J1614$-$5402 & $0.2811$ & $0.2684$ & $-$ & $-$ & $-$ &  $-$ & $-$ & $-$ &  $0.573$ &  $-$ &  $300.0$ &  $20.78$ &  $0.85$ & $5.83^{+1.17}_{-1.17}$ & $4.64^{+0.86}_{-0.95}$ \\
J1624$-$4411 & $0.2931$ & $0.2513$ & $0.94(49)$ & $-$ & $-3.43(51)$ &  $-7.68(36)$ &  $-$ &  $-$ &  $0.233$ & $-$ &  $139.4$ &  $19.88$ &  $0.95$ & $2.75^{+4.54}_{-1.76}$ & $6.23^{+4.54}_{-3.16}$ \\
J1624$-$4721 & $0.3089$ & $0.2744$ & $0.23(98)$ & $-$ & $-3.8(13)$ &  $-2.46(77)$ &  $-$ &  $-$ &  $0.449$ & $-$ &  $364.0$ &  $20.39$ &  $0.92$ & $3.72^{+4.84}_{-2.40}$ & $3.53^{+7.04}_{-4.85}$ \\
J1638$-$4608 & $0.0562$ & $0.2504$ & $0.39(52)$ & $-$ & $-4.4(10)$ &  $-6.76(84)$ &  $-$ &  $-$ &  $0.278$ & $-$ &  $423.1$ &  $19.38$ &  $0.94$ & $3.65^{+4.74}_{-2.23}$ & $1.98^{+7.76}_{-4.84}$ \\
J1838$-$0549 & $0.5053$ & $0.2536$ & $0.47(0.13)$ & $-$ & $1.64(13)$ &  $1.32(12)$ &  $-$ &  $-$ &  $0.235$ & $-$&  $276.6$ &  $17.70$ &  $0.98$ & $2.54^{+2.25}_{-0.89}$ & $1.18^{+2.44}_{-1.14}$ \\
B1848+12 & $0.3530$ & $0.2503$ & $0.96(1.08)$ & $-$ & $-3.56(1.19)$ &  $-7.61(1.25)$ &  $-$ &  $-$ &  $1.205$ & $-$ &  $70.6$ &  $20.50$ &  $0.97$ & $3.43^{+4.83}_{-2.34}$ & $6.06^{+2.00}_{-3.37}$ \\
J1852+0040 & $0.5057$ & $0.2565$ & $2.39(90)$ &- & $-1.00(76)$ &  $-4.62(66)$ &  $-$ &  $-$ &  $0.105$ & $-$ & $-$ &  $20.21$ &  $0.98$ & $2.15^{+4.77}_{-1.65}$ & $4.39^{+4.74}_{-3.95}$ \\
J1903$-$0258 & $0.3872$ & $0.2505$ & $0.61(0.26)$ & $-$ & $-0.59(0.35)$ &  $3.78(0.34)$ &  $-$ &  $-$ &  $0.301$ & $-$ &  $113.0$ &  $18.93$ &  $0.96$ & $2.69^{+3.93}_{-1.40}$ & $3.70^{+2.38}_{-4.52}$ \\
J1958+2846 & $0.3890$ & $0.2536$ & $0.38(26)$ & $-$ & $-3.64(22)$ &  $-7.26(28)$ &  $-$ &  $-$ &  $0.290$ & $-$ & $-$ &  $19.34$ &  $0.98$ & $3.57^{+4.38}_{-1.89}$ & $2.82^{+2.76}_{-3.38}$ \\
J2027+4557 & $0.2902$ & $0.2500$ & $0.52(3)$ &- & $2.08(3)$ &  $-2.12(3)$ &  $-$ &  $-$ &  $1.100$ & $-$ &  $229.6$ &  $15.71$ &  $0.98$ & $1.94^{+0.18}_{-0.15}$ & $1.64^{+0.28}_{-0.43}$ \\
\enddata
\tablecomments{see Antoniadis (2021) for details}
\end{deluxetable}
\end{longrotatetable}
\acknowledgments

This research project was partly supported by the Stavros Niarchos Foundation (SNF) and the Hellenic Foundation for Research and Innovation (H.F.R.I.) under the 2nd Call of “Science and Society” Action Always strive for excellence – Theodoros Papazoglou” (Project Number: 72-1/11.8.2020).  This work relies on data from the
European Space Agency (ESA) mission \gaia\
(\url{https://www.cosmos.esa.int/gaia}), processed by the
\gaia\ Data Processing and Analysis Consortium (DPAC,
\url{https://www.cosmos.esa.int/web/gaia/dpac/consortium}). 
This research made use of NumPy \citep{2020NumPy-Array}, Matplotlib \citep{Matplotlib} and  Astropy (\url{http://www.astropy.org}), a community-developed core Python package for Astronomy \citep{Robitaille:2013mpa,Price-Whelan:2018hus} and of NASA's Astrophysics Data System Bibliographic Services.

\bibliography{rnaas}

\end{document}